\documentclass{article}
\usepackage[preprint, nonatbib]{neurips_2024}

\usepackage[utf8]{inputenc} 
\usepackage[T1]{fontenc}    
\usepackage{hyperref}       
\usepackage{url}            
\usepackage{booktabs}       
\usepackage{amsfonts}       
\usepackage{nicefrac}       
\usepackage{microtype}      
\usepackage{xcolor}         
\usepackage{graphicx}
\usepackage{listings}
\title{GSound-SIR: A Spatial Impulse Response Ray-Tracing and High-order Ambisonic Auralization Python Toolkit}

\author{%
  Yongyi Zang\\
  Independent Researcher\\
  \texttt{zyy0116@gmail.com} \\
  \And
  Qiuqiang Kong\thanks{Corresponding author.} \\
  Chinese University of Hong Kong \\
  \texttt{qqkong@ee.cuhk.edu.hk}
}

\begin{document}

\maketitle

\begin{abstract} 
Accurate and efficient simulation of room impulse responses is crucial for spatial audio applications. However, existing acoustic ray-tracing tools often operate as black boxes and only output impulse responses (IRs), providing limited access to intermediate data or spatial fidelity. To address those problems, this paper presents GSound-SIR, a novel Python-based toolkit for room acoustics simulation that addresses these limitations. The contribution of this paper includes the follows. First, GSound-SIR provides direct access to up to millions of raw ray data points from simulations, enabling in-depth analysis of sound propagation paths that was not possible with previous solutions. Second, we introduce a tool to convert acoustic rays into high-order Ambisonic impulse response synthesis, capturing spatial audio cues with greater fidelity than standard techniques. Third, to enhance efficiency, the toolkit implements an energy-based filtering algorithm and can export only the top-X or top-X-\% rays. Fourth, we propose to store the simulation results into Parquet formats, facilitating fast data I/O and seamless integration with data analysis workflows. Together, these features make GSound-SIR an advanced, efficient, and modern foundation for room acoustics research, providing researchers and developers with a powerful new tool for spatial audio exploration. We release the library under Apache 2.0 License at \url{https://github.com/yongyizang/GSound-SIR}.
\end{abstract}

\section{Introduction} 
\begin{quote}
\textit{``A craftsman who wishes to do his work well must first sharpen his tools.''} 

\hfill \textit{--- Confucius~\cite{soothill1910analects}}
\end{quote}

Spatial audio research in areas like virtual reality (VR) and gaming often uses spatial impulse responses (SIRs)~\cite{merimaa2005spatial} to model how sound propagates over space and time from an impulsive source. By convolving a signal with an SIR, it is possible to recreate auditory cues specific to the target environment. There are three main methods to derive SIRs, including the image-source method, ray tracing, and the finite-difference time-domain (FDTD) method. The image-source method can be inaccurate for complex geometries, while FDTD simulations are computationally expensive for large spaces. Ray tracing, however, often provides the best balance between computational efficiency and acoustic accuracy~\cite{lakka2018spatial}.

A typical ray-tracing workflow for spatial audio involves three steps. First, the 3D scene is loaded, acoustic properties are assigned to surfaces, and the positions of the source and listener are defined. Next, rays are emitted from the source to gather raw data for each ray path, including arrival time, energy coefficients across frequency bands, and metadata. Finally, the collected ray data is auralized, often using Higher-Order Ambisonics (HOA) or head-related transfer functions (HRTFs), to generate the final room impulse response.

Although numerous open-source libraries implement ray tracing for acoustics, most are designed for end-to-end processing and directly output the final auralization. Notable examples include GSound~\cite{schissler2011gsound}, which uses caching for real-time performance in gaming, and vaRays~\cite{ouellet2021live}, which efficiently supports HOA-based spatial audio rendering. However, these solutions typically do not expose intermediate ray data for inspection or modification, and they couple the simulation and auralization steps in ways that complicate adapting or substituting different rendering algorithms.

In this paper, we introduce \textbf{GSound-SIR}, a Python library that extends and modifies GSound to decouple the ray generation step from the auralization step. We add a module to GSound for directly exporting the raw ray data and implement a standalone auralization module capable of producing up to ninth-order Ambisonics efficiently, although users can substitute their own rendering approaches. A key challenge in exposing raw ray data is the substantial input/output overhead of storing and retrieving large volumes of ray information. To address this, we offer energy-based filtering options that export only a subset of rays (e.g., the top-X or top-X\% by energy), thereby reducing storage demands while preserving accurate energy reconstruction.

To our knowledge, GSound-SIR is the first open-source framework to provide full access to intermediate ray-tracing data for spatial audio research. We release our supplementary code under the Apache 2.0 License (while GSound itself remains under a more restrictive academic license) and adopt Parquet-based data storage~\cite{vohra2016apache} for efficient reading and writing of intermediate results. In the following sections, we describe our modifications to GSound, detail our Python-based interface, and present empirical evaluations demonstrating the feasibility and accuracy of the proposed approach. The library, along with documentation and examples, will be made publicly available upon publication.

\section{Related Work}

Sound propagation simulation has evolved along two main trajectories: numerical and geometric methods. We review key developments in both approaches, with a particular focus on geometric methods that inform our work.

\subsection{Numeric Methods}

Early attempts at sound propagation simulation employed numerical approaches like finite-difference time-domain (FDTD) methods, which, while highly accurate, proved computationally intensive and impractical for real-time applications. Subsequent improvements, such as Adaptive Rectangular Decomposition (ARD)\cite{raghuvanshi2009efficient}, reduced simulation time through environment subdivision but still required significant computational resources. Direct-to-Indirect Acoustic Radiance Transfer\cite{antani2011direct} achieved real-time diffuse reflection simulation by adapting radiosity concepts from computer graphics, though at the cost of extensive preprocessing and static geometry requirements.

\subsection{Geometric Methods}

Geometric approaches have proven more tractable for interactive applications due to their efficient calculations. The image-source method~\cite{allen1979image} provides high accuracy by recursively reflecting sources over scene geometry, but its exponential complexity with reflection depth limits practical application. Beam tracing~\cite{funkhouser2004beam} and frustum tracing~\cite{chandak2008ad} methods improved efficiency by propagating volumes rather than individual rays, though geometric complexity still posed performance challenges. Hybrid approaches like RESound~\cite{taylor2009resound} combined ray and frustum tracing to balance accuracy and performance.

Modern systems have built upon these foundations while introducing novel optimizations. GSound~\cite{schissler2011gsound} demonstrated viable real-time simulation through efficient caching, while EVERTims~\cite{poirier2017evertims} integrated beam tracing with spatial audio rendering. Commercial systems like ODEON~\cite{naylor1993odeon} refined hybrid approaches for professional applications, and more recently, vaRays~\cite{ouellet2021live} has focused on generating spatial room impulse responses in Ambisonic format.

\subsection{Spatial Audio Representation}

The representation of spatial sound fields has evolved significantly since the introduction of Ambisonics. Higher-order implementations and alternative approaches like Directional Audio Coding~\cite{pulkki2007spatial} have improved spatial resolution while maintaining practical efficiency. Very recently, researchers have aimed to reconstruct ambisonic representations from mono or stereo signals~\cite{zang2024ambisonizer}.

\section{Implementations}

We build on GSound for its highly efficient and customizable ray-tracing engine, originally developed for gaming. Unlike traditional source-based methods, GSound employs backward ray tracing from the listener's position. The processing pipeline of GSound consists of three main phases. To construct SIRs instead of IRs, we modified the GSound toolbox as follows. 

\subsection{Ray casting}
First, rays are cast in random spherical directions from the listener and traced through the environment to capture specular reflections up to a user-defined depth. These paths are efficiently managed via a hash table system that prevents duplicates and supports \emph{path caching}, which substantially improves temporal coherence and reduces computational cost. However, path caching primarily reduces duplicates with moving subjects. As our primary concern is with stationary subjects, we disable caching to save space and caching-related computation time.

In simulation, GSound cast both \textit{specular rays} and \textit{diffuse rays}. Specular rays reflect off smooth surfaces in a predictable, mirror-like manner, following the law of reflection. Diffuse rays scatter in multiple directions after interacting with rough, matte surfaces, resulting in softer that contribute to ambient sound. 

\subsection{Path validation}
Path validation in ray tracing refers to the process of determining whether a traced ray follows a physically or computationally valid propagation path between the source and receiver. Path validation ensures that only meaningful rays contribute to the final result while filtering out invalid or irrelevant rays. For specular reflections, it starts at the listener position and traces rays through the reflection points, checking if each segment is unobstructed. The validation process uses small epsilon offsets to avoid precision issues near surfaces. For each path segment, it checks if the ray intersects any geometry before reaching the next reflection point. If any intersection is found, the path is considered invalid and rejected. For diffraction paths, it additionally verifies that points lie within valid diffraction regions defined by edge wedges. The system also handles path validation for direct visibility between source and listener by sampling multiple rays within the solid angle subtended by the source, computing an average visibility factor. Path validation ensures that only physically realizable propagation paths contribute to the final impulse response.


\subsection{Calculate path properties}
Third, GSound computes the acoustic properties of each path, including total propagation distance, arrival direction, frequency-dependent material attenuation, and Doppler effects.


GSound’s implementation is centered around the \texttt{SoundPropagator} class, which manages three main data structures. \texttt{ListenerData} tracks the state of the listener and the simulated impulse response (IR), \texttt{SourceData} handles source properties and caching, and \texttt{ThreadData} facilitates multi-threading through thread-local storage. Each ray is processed using \texttt{FrequencyBandResponse} objects to account for material reflectivity and air absorption across different frequencies. Path data is double-buffered between worker threads and the main thread and is stored in \texttt{DiffusePathData} and \texttt{SpecularPathData} arrays for valid diffuse and specular paths. The entire system operates on a scene graph that contains sources, listeners, and geometry with frequency-dependent material properties.


We implement a \texttt{getPathDataArrays} method that extracts acoustic path data from \texttt{ListenerData} given a \texttt{SourceData}, then allocates several \texttt{ArrayList} containers for path attributes:source indices, path types, distances, 3D vectors for listener and source directions (encoded as x,y,z), relative speeds, speeds of sound, and frequency-band intensities. By iterating over sources and paths, each \texttt{SoundPath} is decomposed into these arrays while preserving consistent indexing for efficient data access.

\subsection{Python Interface}

We also provide a method for processing simulation data for Python integration via pybind11 \texttt{pybind11}~\cite{pybind11}. For each listener, our toolbox retrieves the raw path data arrays using \texttt{getPathDataArrays} and applies an energy-based path filter. Paths are sorted by total energy (summed over all frequency bands), and a subset is retained based on either an energy percentage threshold (top-X-\%) or a maximum path count (top-X). The filtered paths are then packaged into a Python dictionary containing NumPy~\cite{harris2020array} arrays (\texttt{py::array\_t}), with dimensions \texttt{[num\_paths, 3]} for direction vectors and \texttt{[num\_paths, num\_bands]} for intensities, plus 1D arrays for scalar attributes. A nested dictionary is returned for all listeners. We perform filtering post-generation for the following reasons: First, during development, we found that early thresholding did not significantly accelerate generation, as GSound already includes validation. Second, the primary computation time bottleneck was writing data to disk rather than the generation process itself. Third, post-generation filtering allows for relative criteria (e.g., top-X) instead of a strict limit (e.g., X dB SPL), providing greater flexibility.

\section{High-order Ambisonic (HOA) auralization}

We implement Ambisonic auralization following \cite{sloan2013efficient}, which employs a code generator that leverages recurrence relations and Cartesian coordinates instead of computationally expensive trigonometric functions. Recurrence relations are mathematical equations that define a sequence recursively, meaning each term in the sequence is expressed in terms of one or more previous terms. We extend this approach to 9\textsuperscript{th}-order Ambisonic impulse responses using a noise-based process. 


For each ray, we gate a white noise signal using the energy intensities at each frequency band, then evaluate the coefficients for spherical harmonics up to the chosen Ambisonic order. The final impulse responses accumulates over these filtered signals.

To speed up the simulation, we apply single-instruction-multiple-data (SIMD) which is a parallel computing architecture where a single instruction is executed simultaneously on multiple data points. While the implementation is written in C++ for efficiency, pybind11 bindings enable Python-based interaction with the raw ray data, allowing users to modify the data flexibly before auralization. This design supports both high-performance simulation and user-driven customization in Python.

\begin{figure}[t!]
    \centering
    \includegraphics[width=\linewidth]{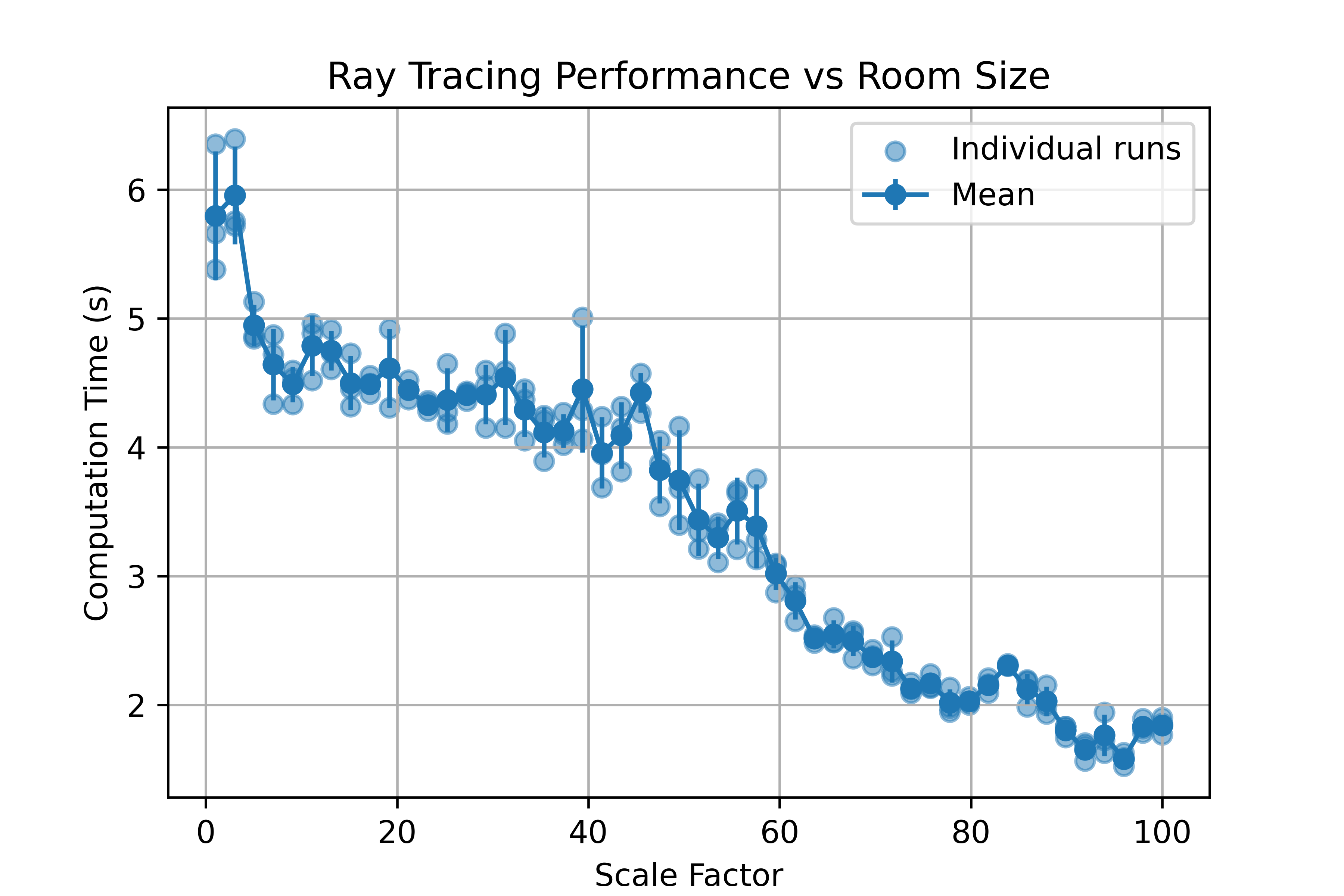}
    \caption{Computation time under various room sizes. Scale factor goes from 0.1 to 10, indicating room sizes from 1 m $\times$ 0.6 m $\times$ 0.4 m to 100 m $\times$ 60 m $\times$ 40 m}
    \label{fig:room_size_benchmark}
\end{figure}

\section{Experiments}
To evaluate GSound-SIR's performance under realistic audio processing conditions, we conducted extensive benchmarking to analyze the impact of various hyperparameters on computational efficiency. All experiments were performed using Python 3.10.13 and GCC 11.2.0 on Ubuntu 22.04, running on an Intel Xeon E5-2620 v2 processor (2.10 GHz, 15M Cache). While the core algorithms are implemented in C++, we measured performance through Python interfaces to simulate typical research workflows. Our measurements cover the entire execution cycle, including all overhead costs associated with Python function calls and handling of results.

While GSound-SIR supports multi-threading, we limited our benchmarks to single-threaded execution to establish baseline performance metrics that could inform multi-core implementations. Initial tests showed that parallel execution across CPU cores caused significant performance degradation, likely due to memory bandwidth limitations at different cache levels. As a result, all benchmarks were conducted sequentially to ensure consistency and reproducibility.

In all experiments, the listener was positioned at coordinates (1.0, 1.0, 0.5) and the sound source at (5.0, 3.0, 0.5). Each measurement was repeated three times to calculate the mean values and standard deviations.

\begin{figure*}[hbt!]
    \centering
    \includegraphics[width=\textwidth]{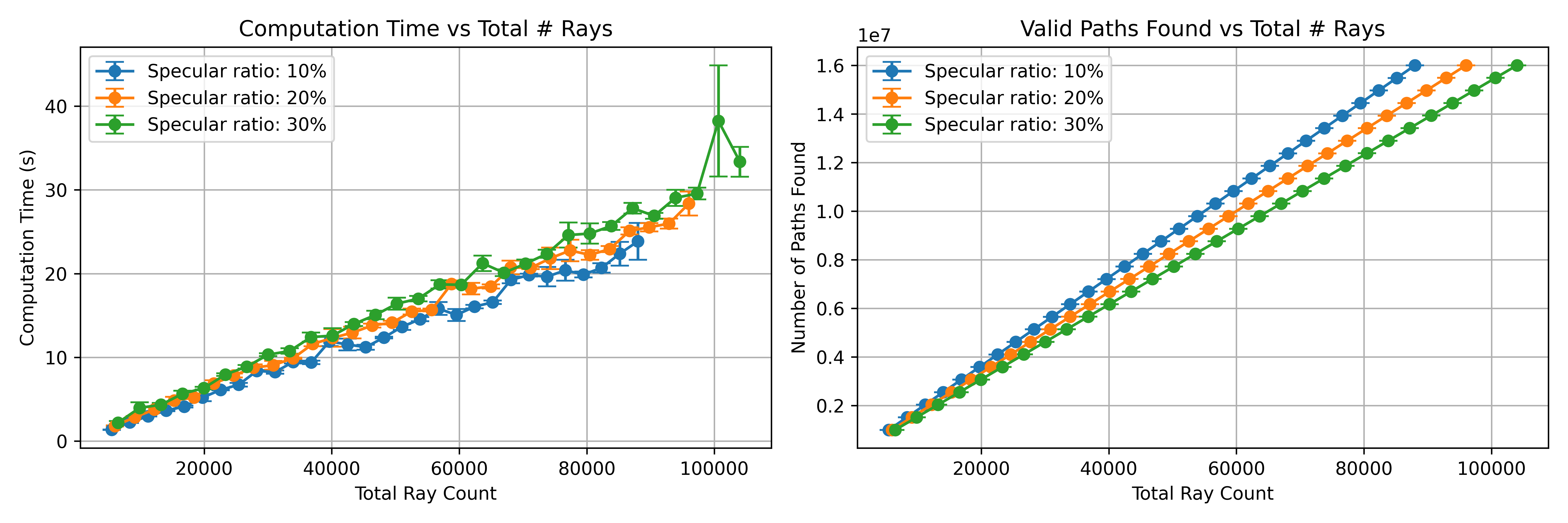}
    \caption{Performance analysis showing the relationships between computation time, total ray count, and successfully identified valid paths. Results are grouped by specular-to-diffuse ray ratios.}
    \label{fig:ray_count_benchmark}
\end{figure*}
\begin{figure*}[t!]
    \centering
    \includegraphics[width=\textwidth]{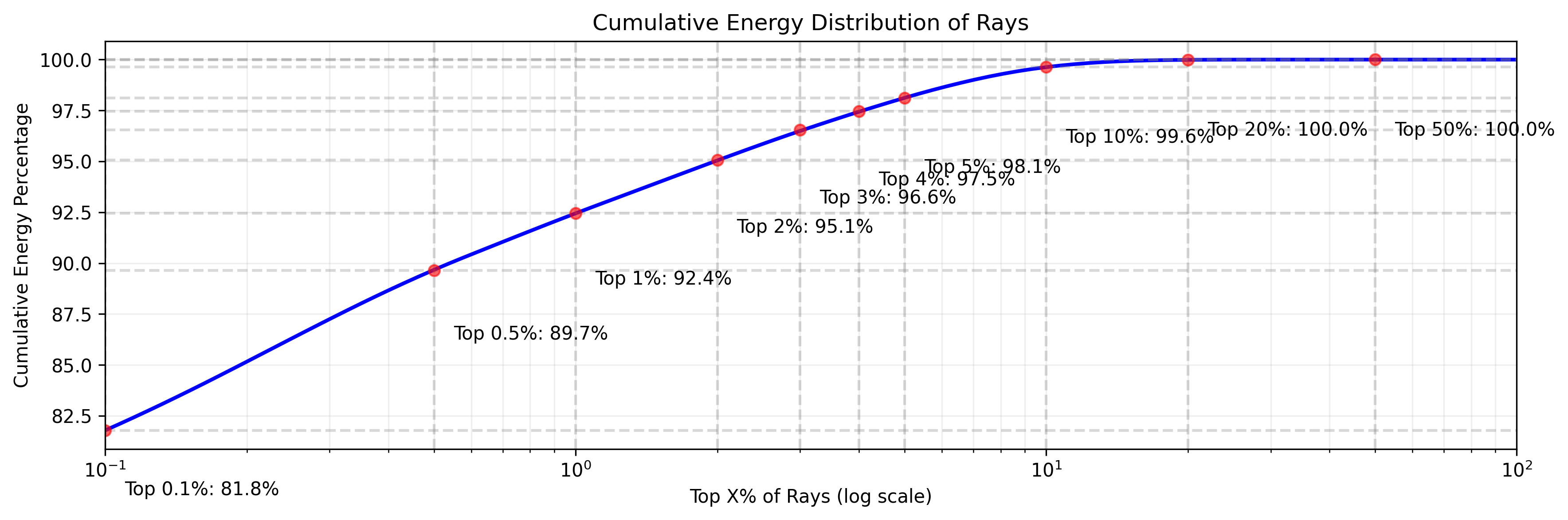}
    \caption{Cumulative acoustic energy distribution showing the total energy captured as a function of the percentage of highest-energy rays included.}
    \label{fig:ray_energy_distribution}
\end{figure*}

\subsection{Room Size}
We evaluated the system's ray-tracing computation time and quality performance across different room sizes by scaling a shoebox-shaped room. The standard room size has dimensions of 10 m in length, 4 m in width, and 4 m in height. We scaled the room size by factors ranging from 0.1 to 10.0, with intervals of 0.1. This resulted in the smallest room size of 1 m $\times$ 0.6 m $\times$ 0.4 m and the largest room size of 100 m $\times$ 60 m $\times$ 40 m. The room's acoustic properties were kept constant, with uniform absorption and scattering coefficients of 0.3 and 0.1, respectively, across all frequency bands. To ensure consistent ray coverage regardless of room size, we fixed the number of rays emitted from the listener position at 20,000 diffuse rays and 2,000 specular rays. Both listener and source positions were scaled proportionally with room dimensions to maintain their relative spatial relationships.

As shown in Figure~\ref{fig:room_size_benchmark}, We observed an inverse relationship between room size and computation time. This counterintuitive result appears to be due to decreased number of valid ray paths in larger rooms, where rays are less likely to find valid reflection paths between the source and listener.

\subsection{Ray Count}
We studied the effect of ray count on performance by varying both diffuse and specular rays while keeping the room size fixed (scaling factor = 1.0). The number of diffuse rays was adjusted from 5,000 to 80,000 in 30 uniform increments. For each diffuse ray count, we tested three different specular ray ratios: 10\%, 20\%, and 30\% of the diffuse ray count, which are commonly used in practical applications.

Figure~\ref{fig:ray_count_benchmark} shows two key findings. First, both computation time and the number of valid paths scale linearly with the total ray count. Second, for the same total ray count, configurations with higher specular ray percentages consistently have longer computation times, suggesting that specular rays involve greater computational overhead than diffuse rays. This difference is likely due to the more complex reflection path validation required for specular rays.

\subsection{Energy Distribution}

We analyzed the energy distribution of 80,000 diffuse rays and 24,000 specular rays (30\% ratio), examining the cumulative energy contribution from all frequency bands. Analysis of the 16,005,213 rays revealed a highly concentrated energy distribution, as shown in Figure~\ref{fig:ray_energy_distribution}: the top 0.1\% (~16,000 rays) account for 81.8\% of the total energy, while the top 10\% capture 99.6\%. This concentration highlights significant computational redundancy in exhaustive ray tracing calculations.

\begin{figure}[t!]
    \centering
    \includegraphics[width=\linewidth]{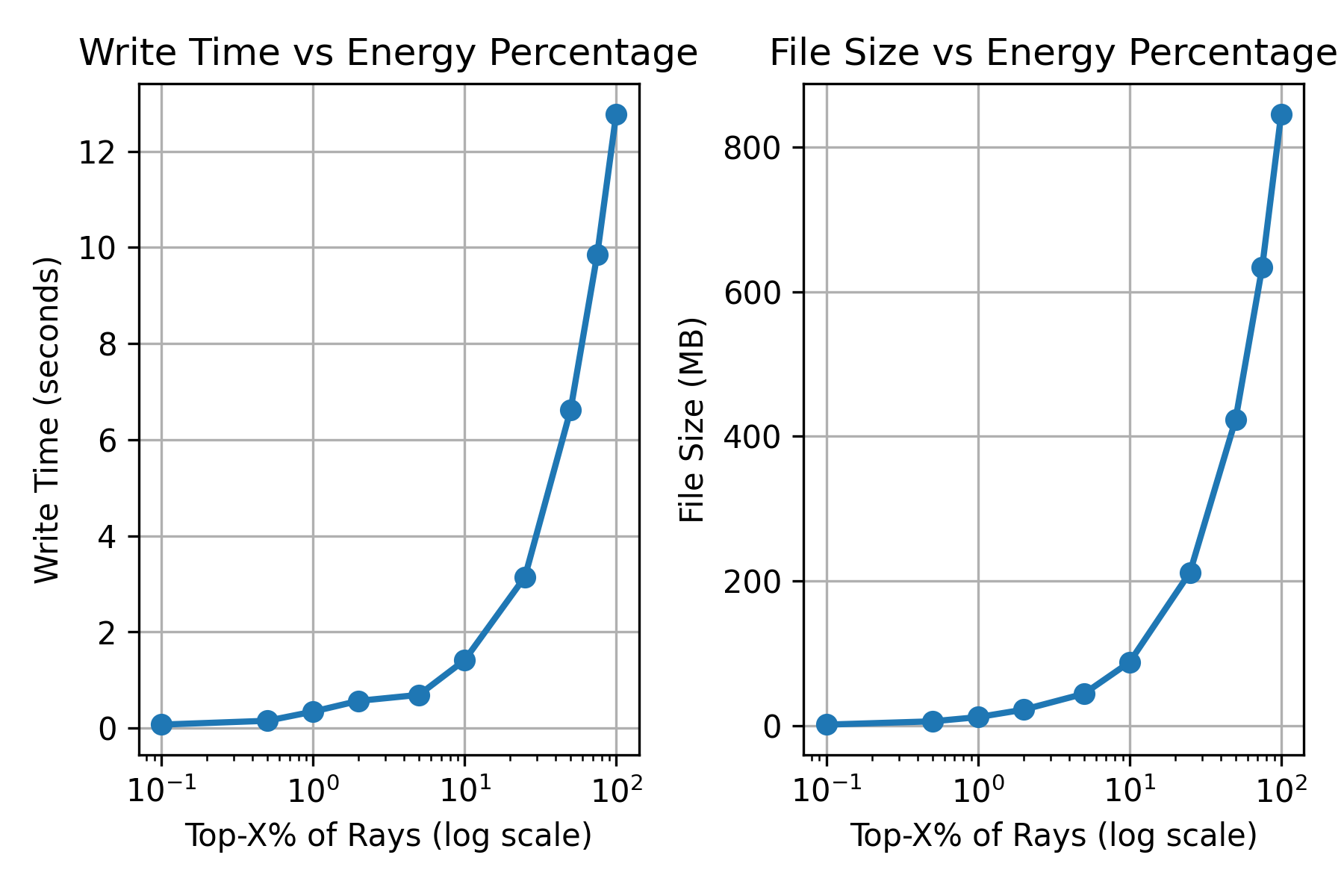}
    \caption{Storage performance metrics showing the relationship between disk write time and resulting file size as a function of the percentage of highest-energy rays stored.}
    \label{fig:disk_write}
\end{figure}

\subsection{Storage Performance Analysis}

Using the same experimental setup, we evaluated storage performance by measuring disk write times and file sizes for different percentages of the highest-energy rays. The results in Figure~\ref{fig:disk_write} show a linear relationship between the percentage of rays stored and both write time and file size. Combined with the energy distribution findings, these results support an optimization strategy of storing only the highest-energy rays, which allows for significant storage savings while retaining most of the acoustic energy information.

\section{Limitations and Future Work}

In this paper, we introduce GSound-SIR, a Python-based toolkit for room acoustics simulation that provides direct access to raw ray data and supports high-order Ambisonic impulse response synthesis. While the toolkit offers significant advantages over existing solutions, it has some limitations. First, it only supports stationary sound sources, limiting its use in dynamic environments or gaming applications. Second, the toolkit lacks GPU acceleration support, which may restrict its performance in computationally intensive scenarios with complex geometries or high ray counts.

Looking ahead, we identify several promising directions for improvement. A key area for expansion is the Ambisonic auralization module. Although functional, the module could be enhanced to support a wider range of filters, potentially unifying different SIR ray auralization methods under a single framework. This would increase flexibility in spatial audio rendering and enable more advanced auralization techniques.

Our analysis of ray energy distribution shows significant redundancy in the generated rays, indicating potential for deep learning-based approaches. Similar to successful applications in computer vision~\cite{watson2020deep} and audio upsampling~\cite{zang2024ambisonizer}, we aim to develop spatial upsamplers that efficiently reconstruct high-quality spatial impulse response data from a reduced set of rays. These methods could greatly improve computational efficiency while preserving acoustic accuracy. We also plan to explore GPU acceleration techniques to boost performance for large-scale simulations and real-time applications.

These improvements would position GSound-SIR as a comprehensive solution for spatial audio research and development, especially in applications that require high-fidelity acoustic simulation with detailed access to propagation path data.

\section{Conclusion}
We introduce GSound-SIR, a Python-based toolkit that enhances acoustic ray-tracing by providing direct access to raw ray data and supporting high-order Ambisonic impulse response synthesis. Our implementation overcomes key limitations of existing solutions by separating ray generation and auralization, allowing researchers to inspect, modify, and optimize intermediate simulation results. The toolkit’s energy-based filtering and efficient Parquet storage system enable effective handling of large ray datasets, while its Python interface supports integration with modern research workflows. Benchmarking shows that the system scales linearly with ray count, and significant computational savings can be achieved through intelligent ray filtering with minimal loss of acoustic accuracy. By releasing the toolkit under the Apache 2.0 License, we aim to advance spatial audio research, providing a flexible and efficient foundation for exploring new auralization techniques and developing next-generation spatial audio applications.
\medskip
\small

\bibliographystyle{plain}
\bibliography{tfg}
\end{document}